\documentclass[11pt]{article}%
\usepackage{amsfonts}
\usepackage{amsmath}
\usepackage{amssymb}
\usepackage{charter}
\usepackage{graphicx}
\usepackage{cite}
\usepackage[onehalfspacing]{setspace}%
\providecommand{\U}[1]{\protect \rule{.1in}{.1in}}

\RequirePackage{CJK}
\AtBeginDocument{\begin{CJK*}{GBK}{song}\CJKtilde}
\AtEndDocument{\end{CJK*}}
\begin{document}

\title{Relativistic spinor equation of photon}
\author{{\small Xiang-Yao Wu}$^{a}${\small \thanks{E-mail: wuxy2066@163.com}, Hong
Li}$^{a}${\small , Xiao-Jing Liu}$^{a}${\small , Si-Qi Zhang}$^{a}${\small , }
\and {\small Ji Ma}$^{a}${\small , Guang-Huai Wang}$^{a}${\small , Hai-Xin
Gao}$^{b}${\small , Heng-Mei Li}$^{c}${\small and Hong-Chun Yuan}$^{c}$\\{\small a) Institute of Physics, Jilin Normal University, Siping 136000 China}\\{\small b) Institute of Physics, Northeast Normal University, Changchun 130024
China}\\{\small c) College of Optoelectronic Engineering, Changzhou Institute of
Technology, Changzhou 213002, China}}
\maketitle

\begin{abstract}
In this paper, we have proposed the spiron equation of free and non-free
photon, and give the spin operator and spin wave function of photon. We
calculate the helicity of photon and prove there are left-handed and
right-handed photon. By the spiron equation of non-free photon, we can study
the quantum property of photon in medium, which can be used in quantum optics,
photonic crystals and so on. \newline



\end{abstract}

\section{Introduction}

Maxwell has unified the laws of electricity and magnetism in a consistent way
into a set of four equations, and he calculated the speed of electromagnetic
waves and found that it was $300,000km/sec$, which was the same of light. This
forced Maxwell to ponder about the nature of light: he concluded that light is
a form of electromagnetic wave. For the explanation the spectra of black-body
radiation, Planck firstly proposed that emission of black-body was energy
quantization with value of $\hbar \omega$, and he introduced the Planck
constant $\hbar$ \cite{s1}. In photoelectric effect experiment, Einstein only
considered the energy conservation and made use of Planck's $\hbar \omega$ to
propose a quanta concept of light \cite{s2}, and successfully explaining
photoelectric effect phenomena, It was the beginning of Quantum Physics.

Despite claims that massless particles cannot be localized in space-time
\cite{s3,s4}, physicists have pondered over the problem of photon localization
and the related problem of the photon wave function for almost $80$ years now,
beginning with the work of Landau and Peierls \cite{s5}. An extensive review
of the photon localization problem was presented by Keller \cite{s6}. The
problem of photon localization is closely related to the widely studied
problem of the photon position operator. Quantum optics has long struggled to
define a quantum wavefunction for photons that would be in agreement with
innumerable experiments revealing the possibility of photon localization. The
photon wave function appears in both of these descriptions and that it
provides a very convenient concept to unify the two points of view. A very
useful mathematical tool in this analysis is the Riemann-Silberstein vector
\cite{s7,s8,s9,s10,s11}. Applications of the $RS$ vector to many physical
problems were reviewed in a very thorough paper by Keller \cite{s6}.

After Dirac discovered the relativistic equation for a particle with spin 1/2
\cite{s12}, much work was done to study spinor and vectors within the Lorentz
group theory. Any quantity which transforms linearly under Lorentz
transformations is a spinor. For that reason spinor quantities are considered
as fundamental in quantum field theory and basic equations for such quantities
should be written in a spinor form. A spinor formulation of Maxwell equations
was studied by Laporte and Uhlenbeck \cite{s13}, also see Rumer \cite{s14}. In
1931, Oppenheimer \cite{s15} proposed to consider the Maxwell theory of
electromagnetism as the wave mechanics of the photon. They introduced a
complex 3-vector wave function satisfying the massless Dirac-like equations.
In this paper, we shall give the relativistic spinor equation for photon with
spin $1$ and mass $0$.

\section{Relativistic spinor equation of free photon}

Dirac derived the Dirac equation by factorizing Einstein's dispersion relation
such that the field equation becomes the first order in time derivative
\cite{s16}. Namely, he factorized the relativistic dispersion relation
employing four by four matrices%

\begin{equation}
E^{2}-c^{2}\vec{p}\hspace{0.05in}^{2}-m_{0}^{2}c^{4}=(E-c\vec{p}\cdot
\vec{\alpha}-m_{0}c^{2}\beta)(E+c\vec{p}\cdot \vec{\alpha}+m_{0}c^{2}%
\beta),\label{1}%
\end{equation}
where $\vec{\alpha}$ and $\beta$ are Dirac matrices. For a photon, mass
$m_{0}=0$, equation (\ref{1}) becomes
\begin{equation}
E^{2}-c^{2}\vec{p}\hspace{0.05in}^{2}=(E-c\vec{\alpha}\cdot \vec{p}%
)(E+c\vec{\alpha}\cdot \vec{p}),\label{2}%
\end{equation}
i.e.,
\begin{equation}
E-c\vec{\alpha}\cdot \vec{p}=0,\label{3}%
\end{equation}
canonical quantization equation (\ref{3}), we have
\begin{equation}
i\hbar \frac{\partial}{\partial t}\psi=-ic\hbar \vec{\alpha}\cdot \vec
{\bigtriangledown}\psi=H\psi,\label{4}%
\end{equation}
where the photon's Hamiltonian operator $H=-ic\hbar \vec{\alpha}\cdot
\vec{\bigtriangledown}$, and $\psi$ is spiron wavefunction of photon. For the
proper Lorentz group $L_{p}$, the irreducibility representations of spin $1$
photon is $D^{10}$, $D^{01}$ and $D^{\frac{1}{2}\frac{1}{2}}$, respectively,
and the dimension of irreducibility representations are corresponding to
three, three and four. We choose photon's spiron wave function as the basis
vector of three dimension irreducibility representations, it is
\begin{equation}
\psi=\left(
\begin{array}
[c]{c}%
\psi_{1}\\
\psi_{2}\\
\psi_{3}%
\end{array}
\right)  ,\label{5}%
\end{equation}
and $\vec{\alpha}$ matrices are taken as
\begin{equation}
\alpha_{x}=\left(
\begin{array}
[c]{ccc}%
0 & 0 & 0\\
0 & 0 & -i\hbar \\
0 & i\hbar & 0
\end{array}
\right)  ,\alpha_{y}=\left(
\begin{array}
[c]{ccc}%
0 & 0 & i\hbar \\
0 & 0 & 0\\
-i\hbar & 0 & 0
\end{array}
\right)  ,\alpha_{z}=\left(
\begin{array}
[c]{ccc}%
0 & -i\hbar & 0\\
i\hbar & 0 & 0\\
0 & 0 & 0
\end{array}
\right)  ,\label{6}%
\end{equation}
they are Hermitian matrices
\begin{equation}
\vec{\alpha}^{\dag}=\vec{\alpha},\label{7}%
\end{equation}
and photon's Hamiltonian operator is Hermitian
\begin{equation}
{H}^{\dag}={H}.\label{8}%
\end{equation}

\section{The spin operators of photon}

In the following, we shall prove the selection of $\vec{\alpha}$ matrices are
reasonable, they confirm that equation (4) is photon's spiron equation of spin
$1$. equation (\ref{4}) can be written as
\begin{equation}
i\hbar \frac{\partial}{\partial t}\psi=c(\vec{p}\cdot \vec{\alpha})={H\psi
},\label{9}%
\end{equation}
where
\begin{equation}
{H}=c\vec{p}\cdot \vec{\alpha},\label{10}%
\end{equation}
The orbital angular momentum of photon satisfy
\begin{align}
\frac{d}{dt}{L_{x}} &  =\frac{1}{i\hbar}[{L_{x}},{H}]\nonumber \\
&  =c(\alpha_{y}{p_{z}}-\alpha_{z}{p_{y}})\nonumber \\
&  =c(\vec{\alpha}\times \vec{p})_{x},\label{11}%
\end{align}
i.e.,
\begin{equation}
\lbrack \vec{L},{H}]=i\hbar c(\vec{\alpha}\times \vec{p}),\label{12}%
\end{equation}
the equation (\ref{12}) is shown that the orbital angular momentum of photon
isn't conservation, but the total angular momentum of free photon should be
conservative. So, photon should have an intrinsic angular momentum, i.e., spin
angular momentum $\vec{s}$, the total angular momentum of photon $\vec{J}$ is
\begin{equation}
{\vec{J}}={\vec{L}}+{\vec{s}},\label{13}%
\end{equation}
and $\vec{J}$ is conservative
\begin{equation}
\lbrack{\vec{J}},{H}]=0,\label{14}%
\end{equation}
with equations (\ref{12}) and (\ref{14}), we have
\begin{equation}
\lbrack \vec{s},{H}]=-[{\vec{L}},{H}]=-i\hbar c(\vec{\alpha}\times \vec
{p}),\label{15}%
\end{equation}
i.e.,
\begin{align}
\lbrack{s}_{x},{H}] &  =[{s}_{x},c\vec{\alpha}\cdot \vec{p}]=-i\hbar
c(\vec{\alpha}\times \vec{p})_{x}\nonumber \\
&  =i\hbar c(\alpha_{z}{p}_{y}-\alpha_{y}{p}_{z}),\label{16}%
\end{align}
or
\begin{align}
\lbrack{s}_{x},c\alpha_{x}p_{x}+c\alpha_{y}p_{y}+c\alpha_{z}p_{z}] &
=c[s_{x},\alpha_{x}]p_{x}+c[s_{x},\alpha_{y}]p_{y}+c[s_{x},\alpha_{z}%
]p_{z}\nonumber \\
&  =i\hbar c(\alpha_{z}{p}_{y}-\alpha_{y}{p}_{z}),\label{17}%
\end{align}
there are commutation relations
\begin{equation}
\lbrack s_{x},\alpha_{x}]=0,\hspace{0.2in}[s_{x},\alpha_{y}]=i\hbar \alpha
_{z},\hspace{0.2in}[s_{x},\alpha_{z}]=-i\hbar \alpha_{y},\label{18}%
\end{equation}
it is easy to summarize that
\begin{equation}
\lbrack s_{i},\alpha_{j}]=i\hbar \epsilon_{ijk}\alpha_{k},\label{19}%
\end{equation}
with equation (\ref{18}), we can calculate the $s$ matrices, let the $s_{x}$
matrix is
\begin{equation}
s_{x}=\left(
\begin{array}
[c]{ccc}%
a_{11} & a_{12} & a_{13}\\
a_{21} & a_{22} & a_{23}\\
a_{31} & a_{32} & a_{33}%
\end{array}
\right)  ,\label{20}%
\end{equation}
with the commutation relation
\begin{equation}
\lbrack s_{x},\alpha_{y}]=i\hbar \alpha_{z},\hspace{0.2in}[s_{x},\alpha
_{z}]=-i\hbar \alpha_{y},\label{21}%
\end{equation}
we get
\begin{equation}
s_{x}=\left(
\begin{array}
[c]{ccc}%
a & 0 & 0\\
0 & a & -i\hbar \\
0 & i\hbar & a
\end{array}
\right)  ,\label{22}%
\end{equation}
let the $s_{y}$ matrix is
\begin{equation}
s_{y}=\left(
\begin{array}
[c]{ccc}%
b_{11} & b_{12} & b_{13}\\
b_{21} & b_{22} & b_{23}\\
b_{31} & b_{32} & b_{33}%
\end{array}
\right)  ,\label{23}%
\end{equation}
with the commutation relation
\begin{equation}
\lbrack s_{y},\alpha_{x}]=-i\hbar \alpha_{z},\hspace{0.2in}[s_{y},\alpha
_{z}]=i\hbar \alpha_{x},\label{24}%
\end{equation}
we get
\begin{equation}
s_{y}=\left(
\begin{array}
[c]{ccc}%
b & 0 & i\hbar \\
0 & b & 0\\
-i\hbar & 0 & b
\end{array}
\right)  ,\label{25}%
\end{equation}
let the $s_{z}$ matrix is
\begin{equation}
s_{z}=\left(
\begin{array}
[c]{ccc}%
c_{11} & c_{12} & c_{13}\\
c_{21} & c_{22} & c_{23}\\
c_{31} & c_{32} & c_{33}%
\end{array}
\right)  ,\label{26}%
\end{equation}
with the commutation relation
\begin{equation}
\lbrack s_{z},\alpha_{x}]=i\hbar \alpha_{y},\hspace{0.2in}[s_{z},\alpha
_{y}]=-i\hbar \alpha_{x},\label{27}%
\end{equation}
we get
\begin{equation}
s_{z}=\left(
\begin{array}
[c]{ccc}%
c & -i\hbar & 0\\
i\hbar & c & 0\\
0 & 0 & c
\end{array}
\right)  .\label{28}%
\end{equation}

In the following, we should calculate the eigenvalues of $s_{x}$, $s_{y}$,
$s_{z}$.\newline The $s_{x}$ eigenvalue problem $s_{x}\psi=\lambda_{1}\psi$
is
\begin{equation}
\left(
\begin{array}
[c]{ccc}%
a & 0 & 0\\
0 & a & -i\hbar \\
0 & i\hbar & a
\end{array}
\right)  \left(
\begin{array}
[c]{c}%
\psi_{1}\\
\psi_{2}\\
\psi_{3}%
\end{array}
\right)  =\lambda_{1}\left(
\begin{array}
[c]{c}%
\psi_{1}\\
\psi_{2}\\
\psi_{3}%
\end{array}
\right)  ,\label{29}%
\end{equation}
therefore the characteristic equation is
\begin{equation}
\left \vert
\begin{array}
[c]{ccc}%
a-\lambda_{1} & 0 & 0\\
0 & a-\lambda_{1} & -i\hbar \\
0 & i\hbar & a-\lambda_{1}%
\end{array}
\right \vert =0,\label{30}%
\end{equation}
i.e.,
\begin{equation}
(a-\lambda_{1})[(a-\lambda_{1})^{2}-\hbar^{2}]=0,\label{31}%
\end{equation}
when $a=0$, the roots $\lambda_{1}$ are
\begin{equation}
\lambda_{1}=\pm \hbar,\label{32}%
\end{equation}
the $s_{y}$ eigenvalue problem $s_{y}\psi=\lambda_{2}\psi$ is
\begin{equation}
\left(
\begin{array}
[c]{ccc}%
b & 0 & i\hbar \\
0 & b & 0\\
-i\hbar & 0 & b
\end{array}
\right)  \left(
\begin{array}
[c]{c}%
\psi_{1}\\
\psi_{2}\\
\psi_{3}%
\end{array}
\right)  =\lambda_{2}\left(
\begin{array}
[c]{c}%
\psi_{1}\\
\psi_{2}\\
\psi_{3}%
\end{array}
\right)  ,\label{33}%
\end{equation}
therefore the characteristic equation is
\begin{equation}
\left \vert
\begin{array}
[c]{ccc}%
b-\lambda_{2} & 0 & i\hbar \\
0 & b-\lambda_{2} & 0\\
-i\hbar & 0 & b-\lambda_{2}%
\end{array}
\right \vert =0,\label{34}%
\end{equation}
i.e.,
\begin{equation}
(b-\lambda_{2})[(b-\lambda_{2})^{2}-\hbar^{2}]=0,\label{35}%
\end{equation}
when $b=0$, the roots $\lambda_{2}$ are
\begin{equation}
\lambda_{2}=\pm \hbar,\label{36}%
\end{equation}
the $s_{z}$ eigenvalue problem $s_{z}\psi=\lambda_{3}\psi$ is
\begin{equation}
\left(
\begin{array}
[c]{ccc}%
c & -i\hbar & 0\\
i\hbar & c & 0\\
0 & 0 & c
\end{array}
\right)  \left(
\begin{array}
[c]{c}%
\psi_{1}\\
\psi_{2}\\
\psi_{3}%
\end{array}
\right)  =\lambda_{3}\left(
\begin{array}
[c]{c}%
\psi_{1}\\
\psi_{2}\\
\psi_{3}%
\end{array}
\right)  ,\label{37}%
\end{equation}
therefore the characteristic equation is
\begin{equation}
\left \vert
\begin{array}
[c]{ccc}%
c-\lambda_{3} & -i\hbar & 0\\
i\hbar & c-\lambda_{3} & 0\\
0 & 0 & c-\lambda_{3}%
\end{array}
\right \vert =0,\label{38}%
\end{equation}
i.e.,
\begin{equation}
(c-\lambda_{3})[(c-\lambda_{3})^{2}-\hbar^{2}]=0,\label{39}%
\end{equation}
when $c=0$, the roots $\lambda_{3}$ are
\begin{equation}
\lambda_{3}=\pm \hbar.\label{40}%
\end{equation}
By calculation, we obtain the spin matrices of photon, they are
\begin{equation}
s_{x}=\left(
\begin{array}
[c]{ccc}%
0 & 0 & 0\\
0 & 0 & -i\hbar \\
0 & i\hbar & 0
\end{array}
\right)  ,s_{y}=\left(
\begin{array}
[c]{ccc}%
0 & 0 & i\hbar \\
0 & 0 & 0\\
-i\hbar & 0 & 0
\end{array}
\right)  ,s_{z}=\left(
\begin{array}
[c]{ccc}%
0 & -i\hbar & 0\\
i\hbar & 0 & 0\\
0 & 0 & 0
\end{array}
\right)  .\label{41}%
\end{equation}
Obviously, these matrices are photon's spin matrices, which describe the spin
$s=1$. On the one hand, their eigenvalues are $\pm \hbar$. On the other hand,
the matrices square is
\begin{equation}
{\vec{s}}\hspace{0.05in}^{2}=s_{x}^{2}+s_{y}^{2}+s_{z}^{2}=2\left(
\begin{array}
[c]{ccc}%
1 & 0 & 0\\
0 & 1 & 0\\
0 & 0 & 1
\end{array}
\right)  \hbar^{2}=s(s+1)\hbar^{2}\left(
\begin{array}
[c]{ccc}%
1 & 0 & 0\\
0 & 1 & 0\\
0 & 0 & 1
\end{array}
\right)  ,\label{42}%
\end{equation}
i.e., the spin $s$ is
\begin{equation}
s=1.\label{43}%
\end{equation}
Comparing (\ref{6}) with (\ref{41}), we find
\begin{equation}
s_{x}=\alpha_{x},\hspace{0.2in}s_{y}=\alpha_{y},\hspace{0.2in}s_{z}=\alpha
_{z}.\label{44}%
\end{equation}

\section{The helicity of photon}

By studying the helicity of photon, we can obtain the photon left-handed and
right-handed. The helicity is defined as the projection of spin in the
momentum direction, it is
\begin{equation}
h=\frac{\vec{s}\cdot \vec{p}}{|\vec{p}|}=\frac{\vec{\alpha}\cdot \vec{p}}%
{|\vec{p}|},\label{45}%
\end{equation}
and
\begin{equation}
\vec{\alpha}\cdot \vec{p}=\alpha_{x}p_{x}+\alpha_{y}p_{y}+\alpha_{z}%
p_{z}=\left(
\begin{array}
[c]{ccc}%
0 & -ip_{z} & ip_{y}\\
ip_{z} & 0 & -ip_{x}\\
-ip_{y} & ip_{x} & 0
\end{array}
\right)  ,\label{46}%
\end{equation}
the $\vec{\alpha}\cdot \vec{p}$ eigenvalue problem $\vec{\alpha}\cdot \vec
{p}\psi=\lambda \psi$ is
\begin{equation}
\left(
\begin{array}
[c]{ccc}%
0 & -ip_{z} & ip_{y}\\
ip_{z} & 0 & -ip_{x}\\
-ip_{y} & ip_{x} & 0
\end{array}
\right)  \left(
\begin{array}
[c]{c}%
\psi_{1}\\
\psi_{2}\\
\psi_{3}%
\end{array}
\right)  =\lambda \left(
\begin{array}
[c]{c}%
\psi_{1}\\
\psi_{2}\\
\psi_{3}%
\end{array}
\right)  ,\label{47}%
\end{equation}
therefore the characteristic equation is
\begin{equation}
\left \vert
\begin{array}
[c]{ccc}%
-\lambda & -ip_{z} & ip_{y}\\
ip_{z} & -\lambda & -ip_{x}\\
-ip_{y} & ip_{x} & -\lambda
\end{array}
\right \vert =0,\label{48}%
\end{equation}
i.e.,
\begin{equation}
\lambda(\bar{p}^{2}-\lambda^{2})=0,\label{49}%
\end{equation}
the roots $\lambda$ are
\begin{equation}
\lambda=|\vec{p}|,\hspace{0.15in}-|\vec{p}|,\label{50}%
\end{equation}
and the helicity $h$ are
\begin{equation}
\hbar=+1,\hspace{0.15in}-1.\label{51}%
\end{equation}
When $\lambda=+1$ the photon is called right-handed photon, and when
$\lambda=-1$ the photon is called left-handed photon.

\section{The probability conservation equation of photon}

In the following, we should give the probability density and probability
conservation equation of photon.\newline The hermitian conjugate of (\ref{4})
is
\begin{equation}
-i\hbar \frac{\partial \psi^{\dag}}{\partial t}=i\hbar c\vec{\nabla}\psi^{\dag
}\cdot \vec{\alpha},\label{52}%
\end{equation}
multiplying (\ref{52}) by $\psi$, there is
\begin{equation}
-i\hbar \frac{\partial \psi^{\dag}}{\partial t}\psi=i\hbar c\vec{\nabla}%
\psi^{\dag}\cdot \vec{\alpha}\psi,\label{53}%
\end{equation}
multiplying (\ref{4}) by $\psi^{+}$, there is
\begin{equation}
i\hbar \psi^{\dag}(\frac{\partial \psi}{\partial t})=-i\hbar c\psi^{\dag}%
\vec{\alpha}\cdot \vec{\nabla}\psi,\label{54}%
\end{equation}
taking the difference, we get
\begin{equation}
i\hbar(\psi^{\dag}\frac{\partial \psi}{\partial t}+\frac{\partial \psi^{\dag}%
}{\partial t}\psi)+i\hbar c\psi^{\dag}\vec{\alpha}\cdot(\vec{\nabla}%
\psi)+i\hbar c\vec{\nabla}\psi^{\dag}\cdot \vec{\alpha}\psi=0,\label{55}%
\end{equation}
or
\begin{equation}
\frac{1}{c}\frac{\partial}{\partial t}(\psi^{\dag}\psi)+\psi^{\dag}\vec
{\alpha}\cdot(\vec{\nabla}\psi)+(\vec{\nabla}\psi^{\dag})\cdot \vec{\alpha}%
\psi=0,\label{56}%
\end{equation}
i.e.,
\begin{equation}
\frac{\partial \rho}{\partial t}+\nabla \cdot \vec{J}=0,\label{57}%
\end{equation}
where
\begin{equation}
\rho=\psi^{\dag}\psi,\hspace{0.25in}\vec{J}=c\psi^{\dag}\vec{\alpha}%
\psi,\label{58}%
\end{equation}
are the probability and probability current density of photon, respectively.

\section{The plane wave of free photon}

We have the spiron equation of free photon
\begin{equation}
i\hbar \frac{\partial}{\partial t}\psi=H\psi,\label{59}%
\end{equation}
where
\begin{equation}
H=c\hspace{0.02in}\vec{{\alpha}}\cdot \vec{p},\psi=\left(
\begin{array}
[c]{c}%
\psi_{1}\\
\psi_{2}\\
\psi_{3}%
\end{array}
\right)  ,\label{60}%
\end{equation}
since $\frac{\partial H}{\partial t}=0$ and $[\vec{p},H]=0$, the energy $E$
and momentum $\vec{p}$ of photon are conserved quantity. Their common
eigenstate is
\begin{equation}
\psi_{E,\vec{p}}(\vec{r},t)=u(\vec{p})e^{i(\vec{p}\cdot \vec{r}-Et)/\hbar
},\label{61}%
\end{equation}
where
\begin{equation}
u(\vec{p})=\left(
\begin{array}
[c]{c}%
u_{1}(\vec{p})\\
u_{2}(\vec{p})\\
u_{3}(\vec{p})
\end{array}
\right)  ,\label{62}%
\end{equation}
substituting equations (\ref{61}) and (\ref{62}) into (\ref{59}), we have
\begin{equation}
c\hspace{0.02in}\vec{{\alpha}}\cdot \vec{p}\hspace{0.1in}u(\vec{p}%
)=E\hspace{0.02in}u(\vec{p}),\label{63}%
\end{equation}
i.e.,
\begin{equation}
\left(  c\hspace{0.01in}\alpha_{x}p_{x}+c\hspace{0.01in}\alpha_{y}%
p_{y}+c\hspace{0.01in}\alpha_{z}p_{z}\right)  \left(
\begin{array}
[c]{c}%
u_{1}\\
u_{2}\\
u_{3}%
\end{array}
\right)  =E\left(
\begin{array}
[c]{c}%
u_{1}\\
u_{2}\\
u_{3}%
\end{array}
\right)  ,\label{64}%
\end{equation}
and expanding (\ref{64}), we get
\begin{equation}
\left(
\begin{array}
[c]{ccc}%
0 & -ic\hspace{0.01in}p_{z} & ic\hspace{0.01in}p_{y}\\
ic\hspace{0.01in}p_{z} & 0 & -ic\hspace{0.01in}p_{x}\\
-ic\hspace{0.01in}p_{y} & ic\hspace{0.01in}p_{x} & 0
\end{array}
\right)  \left(
\begin{array}
[c]{c}%
u_{1}\\
u_{2}\\
u_{3}%
\end{array}
\right)  =E\left(
\begin{array}
[c]{c}%
u_{1}\\
u_{2}\\
u_{3}%
\end{array}
\right)  ,\label{65}%
\end{equation}
or
\begin{align}
Eu_{1}+icp_{z}u_{2}-icp_{y}u_{3} &  =0,\label{66}\\
icp_{z}u_{1}-Eu_{2}-icp_{x}u_{3} &  =0,\label{67}\\
icp_{y}u_{1}-icp_{x}u_{2}+Eu_{3} &  =0,\label{68}%
\end{align}
therefore the characteristic equation is
\begin{equation}
\left \vert
\begin{array}
[c]{ccc}%
E & icp_{z} & -icp_{y}\\
icp_{z} & -E & -icp_{x}\\
\hspace{0.05in}icp_{y} & -icp_{x} & E
\end{array}
\right \vert =0,\label{69}%
\end{equation}
the roots $E$ are
\begin{equation}
E_{1}=+c|\vec{p}|,\hspace{0.2in}E_{2}=-c|\vec{p}|.\label{70}%
\end{equation}
From equations (\ref{67}) and (\ref{68}), we have
\begin{align}
icp_{z}p_{y}u_{1}-Ep_{y}u_{2}-icp_{x}p_{y}u_{3} &  =0,\label{71}\\
icp_{y}p_{z}u_{1}-icp_{x}p_{z}u_{2}+Ep_{z}u_{3} &  =0,\label{72}%
\end{align}
taking the difference of equations (\ref{71}) and (\ref{72}), we get
\begin{equation}
(Ep_{y}-icp_{x}p_{z})u_{2}+(Ep_{z}+icp_{x}p_{y})u_{3}=0,\label{73}%
\end{equation}
i.e.,
\begin{equation}
\frac{u_{2}}{u_{3}}=-\frac{Ep_{z}+icp_{x}p_{y}}{Ep_{y}-icp_{x}p_{z}%
},\label{74}%
\end{equation}
substituting (\ref{74}) into (\ref{66}), there is
\begin{equation}
\frac{u_{1}}{u_{3}}=\frac{ic(p_{y}^{2}+p_{z}^{2})}{Ep_{y}-icp_{x}p_{z}%
},\label{75}%
\end{equation}
by (\ref{74}) and (\ref{75}), we have
\begin{equation}
\frac{u_{1}}{u_{2}}=-\frac{ic(p_{y}^{2}+p_{z}^{2})}{Ep_{z}+icp_{x}p_{y}%
},\label{76}%
\end{equation}
the $u(p)$ can be written as
\begin{equation}
u(p)=\left(
\begin{array}
[c]{c}%
u_{1}\\
u_{2}\\
u_{3}%
\end{array}
\right)  =N\left(
\begin{array}
[c]{c}%
ic(p_{y}^{2}+p_{z}^{2})\\
-(Ep_{z}+icp_{x}p_{y})\\
Ep_{y}-icp_{x}p_{z}%
\end{array}
\right)  ,\label{77}%
\end{equation}
where $N$ is normalization constant, it can be obtained by the normalization
\begin{align}
u^{\dag}(p)u(p) &  =N^{2}\left(
\begin{array}
[c]{ccc}%
-ic(p_{y}^{2}+p_{z}^{2}) & -(Ep_{z}-icp_{x}p_{y}) & (Ep_{y}+icp_{x}p_{z})
\end{array}
\right)  \left(
\begin{array}
[c]{c}%
ic(p_{y}^{2}+p_{z}^{2})\\
-(Ep_{z}+icp_{x}p_{y})\\
Ep_{y}-icp_{x}p_{z}%
\end{array}
\right)  \nonumber \\
&  =2E^{2}N^{2}(p_{y}^{2}+p_{z}^{2})=1,\label{78}%
\end{align}
i.e.,
\begin{equation}
N=\sqrt{\frac{1}{2E^{2}(p_{y}^{2}+p_{z}^{2})}},\label{79}%
\end{equation}
then
\begin{equation}
u(p)=\sqrt{\frac{1}{2E^{2}(p_{y}^{2}+p_{z}^{2})}}\left(
\begin{array}
[c]{c}%
ic(p_{y}^{2}+p_{z}^{2})\\
-(Ep_{z}+icp_{x}p_{y})\\
Ep_{y}-icp_{x}p_{z}%
\end{array}
\right)  ,\label{80}%
\end{equation}
The plane wave solution of free photon is
\begin{equation}
\psi_{E,\vec{P}}(\vec{r},t)=\sqrt{\frac{1}{2E^{2}(p_{y}^{2}+p_{z}^{2})}%
}\left(
\begin{array}
[c]{c}%
ic(p_{y}^{2}+p_{z}^{2})\\
-(Ep_{z}+icp_{x}p_{y})\\
Ep_{y}-icp_{x}p_{z}%
\end{array}
\right)  e^{i(\vec{p}\cdot \vec{r}-Et)/\hbar}.\label{81}%
\end{equation}

\section{The spin wave function of photon}

From equations (\ref{41}) and (\ref{42}), we can find $\vec{s}\hspace
{0.05in}^{2}$ commute with $s_{x}$, $s_{y}$ and $s_{z}$, we should calculate
the common eigenstate of $\vec{s}\hspace{0.05in}^{2}$ and $s_{z}$, they are
\begin{equation}
\vec{s}\hspace{0.05in}^{2}\chi_{\mu}=2\hbar^{2}\chi_{\mu},\label{82}%
\end{equation}%
\begin{equation}
s_{z}\chi_{\mu}=\mu \hbar \chi_{\mu},\label{83}%
\end{equation}
where the common eigenstate $(\chi_{\mu})^{T}=(\varphi_{1},\varphi_{2}%
,\varphi_{3})$, the equation (\ref{83}) can be written as
\begin{equation}
\left(
\begin{array}
[c]{ccc}%
0 & -i\hbar & 0\\
i\hbar & 0 & 0\\
0 & 0 & 0
\end{array}
\right)  \left(
\begin{array}
[c]{c}%
\varphi_{1}\\
\varphi_{2}\\
\varphi_{3}%
\end{array}
\right)  =\mu \left(
\begin{array}
[c]{c}%
\varphi_{1}\\
\varphi_{2}\\
\varphi_{3}%
\end{array}
\right)  ,\label{84}%
\end{equation}
therefore the characteristic equation is
\begin{equation}
\left \vert
\begin{array}
[c]{ccc}%
-\mu & -i & 0\\
i & -\mu & 0\\
0 & 0 & -\mu
\end{array}
\right \vert =0,\label{85}%
\end{equation}
i.e.,
\begin{equation}
-\mu(\mu^{2}-1)=0,\label{86}%
\end{equation}
the roots $\mu$ are
\begin{equation}
\mu_{1}=0,\hspace{0.1in}\mu_{2}=1,\hspace{0.1in}\mu_{3}=-1,\label{87}%
\end{equation}
substituting $\mu_{1}=0$ into (\ref{84}), we get
\begin{equation}
\left \{
\begin{array}
[c]{c}%
-i\varphi_{2}=0\\
i\varphi_{1}=0
\end{array}
\right.  ,\label{88}%
\end{equation}
i.e.,
\begin{equation}
\varphi_{1}=\varphi_{2}=0,\hspace{0.2in}\varphi_{3}\neq0,\label{89}%
\end{equation}
the normalization spin wave function is
\begin{equation}
\chi_{0}=\left(
\begin{array}
[c]{c}%
0\\
0\\
1
\end{array}
\right)  ,\label{90}%
\end{equation}
substituting $\mu_{2}=1$ into (\ref{84}), we get
\begin{equation}
\left \{
\begin{array}
[c]{c}%
-i\varphi_{2}=\varphi_{1}\\
i\varphi_{1}=\varphi_{2}\\
\varphi_{3}=0
\end{array}
\right.  ,\label{91}%
\end{equation}
we can take
\begin{equation}
\varphi_{1}=-1,\hspace{0.2in}\varphi_{2}=-i,\hspace{0.2in}\varphi
_{3}=0,\label{92}%
\end{equation}
the normalization spin wave function is
\begin{equation}
\chi_{1}=-\frac{1}{\sqrt{2}}\left(
\begin{array}
[c]{c}%
1\\
i\\
0
\end{array}
\right)  ,\label{93}%
\end{equation}
substituting $\mu_{3}=-1$ into (\ref{84}), we get
\begin{equation}
\left \{
\begin{array}
[c]{c}%
-i\varphi_{2}=-\varphi_{1}\\
i\varphi_{1}=-\varphi_{2}\\
\varphi_{3}=0
\end{array}
\right.  ,\label{94}%
\end{equation}
we can take
\begin{equation}
\varphi_{1}=1,\hspace{0.2in}\varphi_{2}=-i,\hspace{0.2in}\varphi
_{3}=0,\label{95}%
\end{equation}
the normalization spin wave function is
\begin{equation}
\chi_{-1}=\frac{1}{\sqrt{2}}\left(
\begin{array}
[c]{c}%
1\\
-i\\
0
\end{array}
\right)  ,\label{96}%
\end{equation}
and these spin wave functions satisfy the normalization condition
\begin{equation}
\sum_{\alpha}\chi_{\mu}^{\ast}(\alpha)\chi_{\mu^{\prime}}(\alpha)=\delta
_{\mu \mu^{\prime}}.\label{97}%
\end{equation}

\section{The spiron wave equation of non-free photon}

In view of the above, we have given the spiron wave equation of free photon.
In the following, we should give the spiron wave equation of non-free
photon.\newline

For the non-free particle, the Einstein's dispersion relation is
\begin{equation}
(E-V)^{2}=c^{2}\vec{p}\hspace{0.05in}^{2}+m_{0}^{2}c^{4},\label{98}%
\end{equation}
factorizing (\ref{98}), we obtain
\begin{equation}
(E-V)^{2}-c^{2}\vec{p}\hspace{0.05in}^{2}-m_{0}^{2}c^{4}=(E-V-c\vec{p}%
\cdot \vec{\alpha}-m_{0}c^{2}\beta)(E-V+c\vec{p}\cdot \vec{\alpha}+m_{0}%
c^{2}\beta).\label{99}%
\end{equation}
For photon, $m_{0}=0$, equation (\ref{99}) becomes
\begin{equation}
(E-V)^{2}-c^{2}\vec{p}\hspace{0.05in}^{2}=(E-V-c\vec{p}\cdot \vec{\alpha
})(E-V+c\vec{p}\cdot \vec{\alpha})=0,\label{100}%
\end{equation}
or
\begin{equation}
(E-V-c\vec{p}\cdot \vec{\alpha})=0,\label{101}%
\end{equation}
canonical quantization equation (\ref{101}), we have
\begin{equation}
i\hbar \frac{\partial}{\partial t}\psi=-ic\hbar \vec{\alpha}\cdot \vec{\nabla
}\psi+V\psi,\label{102}%
\end{equation}
the potential energy of photon in medium is\cite{s17}
\begin{equation}
V=\hbar \omega(1-n),\label{103}%
\end{equation}
the spiron equation of photon in medium is
\begin{equation}
i\hbar \frac{\partial}{\partial t}\psi=-ic\hbar \vec{\alpha}\cdot \vec{\nabla
}\psi+\hbar \omega(1-n)\psi,\label{104}%
\end{equation}
by the method of separation variable
\begin{equation}
\psi(\vec{r},\vec{t})=\psi(\vec{r})f(t),\label{105}%
\end{equation}
the equation (\ref{104}) becomes
\begin{equation}
i\hbar \frac{\partial}{\partial t}\psi(\vec{r})f(t)=[-ic\hbar \vec{\alpha}%
\cdot \vec{\nabla}+\hbar \omega(1-n)]\psi(\vec{r})f(t),\label{106}%
\end{equation}
or
\begin{equation}
\frac{i\hbar \frac{\partial}{\partial t}f(t)}{f(t)}\psi^{\dag}(\vec{r}%
)\psi(\vec{r})=\psi^{\dag}(\vec{r})[-ic\hbar \vec{\alpha}\cdot \vec{\nabla
}+\hbar \omega(1-n)]\psi(\vec{r}),\label{107}%
\end{equation}
or
\begin{equation}
\frac{i\hbar \frac{\partial}{\partial t}f(t)}{f(t)}=\frac{\psi^{\dag}(\vec
{r})[-ic\hbar \vec{\alpha}\cdot \vec{\nabla}+\hbar \omega(1-n)]\psi(\vec{r}%
)}{\psi^{\dag}(\vec{r})\psi(\vec{r})}=E,\label{108}%
\end{equation}
we have
\begin{equation}
f(t)=f_{0}e^{-iEt/\hbar},\label{109}%
\end{equation}
and
\begin{equation}
\psi^{\dag}(\vec{r})[-ic\hbar \vec{\alpha}\cdot \vec{\nabla}+\hbar
\omega(1-n)]\psi(\vec{r})=E\psi^{\dag}(\vec{r})\psi(\vec{r}),\label{110}%
\end{equation}
i.e.,
\begin{equation}
\lbrack-ic\hbar \vec{\alpha}\cdot \vec{\nabla}+\hbar \omega(1-n)]\psi(\vec
{r})=E\psi(\vec{r}),\label{111}%
\end{equation}
where $E$ is photon total energy. Equation (\ref{111}) is time-dependent
spiron equation of photon, and the general spiron equation of photon in
arbitrary potential energy $V$ is
\begin{equation}
\lbrack-ic\hbar \vec{\alpha}\cdot \vec{\nabla}+V]\psi(\vec{r})=E\psi(\vec
{r}).\label{112}%
\end{equation}

\section{Conclusion}

In this paper, we have proposed the spiron equation of free and non-free
photon, and give the spin operator and spin wave function of photon. We can
further study two and multi-photon spin wave function, and can obtain
multi-photon entanglement state. We calculate the helicity of photon and prove
there are left-handed and right-handed photon. By the spiron equation of
non-free photon, we can study the quantum property of photon in medium. We
think the spiron equation of free and non-free photon can be widely used in
the future.


\begin{thebibliography}{99}                                                                                               %


\bibitem {s1}M. Planck, The Theory of Heat Radiation, Philadel- phia, 1914.

\bibitem {s2}A. Einstein, Concerning a Heuristic Point of View to-ward the
Emission and Transformation of Light, Annals of Physics, Vol. 17, 1905.

\bibitem {s3}T. D. Newton, E.P. Wigner, Rev. Mod. Phys. 21, 400, 1949.

\bibitem {s4}A. S. Wightman, Rev. Mod. Phys. 34, 845, 1962.

\bibitem {s5}L. D. Landau and R. Peierls, Z. Phys. 62, 188, 1930.

\bibitem {s6}O. Keller, Phys. Rep. 411, 1, 2005.

\bibitem {s7}L. Silberstein, Ann. d. Phys. 22, 579, 1907.

\bibitem {s8}H. Bateman H, The Mathematical Analysis of Electrical and Optical
Wave-Motion on the Basis of Maxwells Equations, Cambridge University Press
(1915), reprinted by Dover New York 1955.

\bibitem {s9}J. Stratton, Electromagnetic Theory, McGraw-Hill New York 1941.

\bibitem {s10}I. Bialynicki-Birula, Z. Bialynicka-Birula, Quantum
Electrodynamics, Pergamon, Oxford, 1975.

\bibitem {s11}I. Bialynicki-Birula, Acta Phys. Polon. 86, 97, 1994.

\bibitem {s12}P. Dirac, Proc. Roy. Soc. A117 610, 1928.

\bibitem {s13}O. Laporte, G. Uhlenbeck, Phys. Rev. 37, 1380, 1931.

\bibitem {s14}Yu. Rumer, Spinor Analysis. Moscow. 1936 (in Russian).

\bibitem {s15}J. Oppenheimer, Phys. Rev. 38, 725, 1931.

\bibitem {s16}P. A. M. Dirac, The Principles of Quantum Mechanics, Oxford
University Press, 1958.

\bibitem {s17}Xiang-Yao Wu, Xiao-Jing Liu, and Yi-Heng Wu, et. al., Int J
Theor Phys, 49, 194, 2010.
\end{thebibliography}
\end{document}